\documentclass[aps,prb,twocolumn,showpacs]{revtex4}
\usepackage{graphicx}

\begin{document}

\title{Superconductivity induced by doping platinum in BaFe$_{2}$As$_2$}

\author{Xiyu Zhu$^{1}$, Fei Han$^{1}$, Gang Mu$^{1}$, Peng Cheng$^{1}$, Jun Tang$^{2}$, Jing Ju$^{2}$, Katsumi Tanigaki$^{2}$, and Hai-Hu Wen$^{1}$}\email{hhwen@aphy.iphy.ac.cn }

\affiliation{$^{1}$National Laboratory for Superconductivity,
Institute of Physics and Beijing National Laboratory for Condensed
Matter Physics, Chinese Academy of Sciences, P. O. Box 603, Beijing
100190, China}

\affiliation{$^{2}$World Premier International Research Center,
Tohoku University, Sendai, 980-8578, Japan}
\date{\today}

\begin{abstract}
By substituting Fe with the 5d-transition metal Pt in
BaFe$_2$As$_2$, we have successfully synthesized the superconductors
BaFe$_{2-x}$Pt$_x$As$_2$. The systematic evolution of the lattice
constants indicates that the Fe ions were successfully replaced by
Pt ions. By increasing the doping content of Pt, the
antiferromagnetic order and structural transition of the parent
phase is suppressed and superconductivity emerges at a doping level
of about x = 0.02. At a doping level of x = 0.1, we get a maximum
transition temperature T$_c$ of about 25 K. While even for this
optimally doped sample, the residual resistivity ratio (RRR) is only
about 1.35, indicating a strong impurity scattering effect. We thus
argue that the doping to the Fe-sites naturally leads to a high
level impurity scattering, although the superconductivity can still
survive at about 25 K. The synchrotron powder x-ray diffraction
shows that the resistivity anomaly is in good agreement with the
structural transition. The superconducting transitions at different
magnetic fields were also measured at the doping level of about x =
0.1, yielding a slope of -dH$_{c2}$/dT = 5.4 T/K near T$_c$. Finally
a phase diagram was established for the Pt doped 122 system. Our
results suggest that superconductivity can also be easily induced in
the FeAs family by substituting the Fe with Pt, with almost the
similar maximum transition temperatures as doping Ni, Co, Rh and Ir.

\end{abstract}\pacs{74.70.Dd, 74.25.Fy, 75.30.Fv, 74.10.+v}
\maketitle

\section{Introduction}
The FeAs-based compounds have formed a new family in the field of
high-$T_c$ superconductors\cite{Kamihara2008}. Many new structures
with the FeAs layers have been found, including the so-called 1111
phase (LNFeAsO, AEFeAsF, LN = rare earth elements, AE = alkaline
earth elements)\cite{Kamihara2008,SrFeAsF1,SrFeAsF2}, 122 phase
($AEFe_2As_2$, $AE$ = alkaline earth elements)\cite{Rotter,CWCh},
111 phase (LiFeAs, NaFeAs)\cite{LiFeAs,NaFeAs}, 11 phase
(FeSe)\cite{WuMK}, 32522 phase
(Sr$_3$Sc$_2$O$_5$Fe$_2$As$_2$)\cite{FeAs32522}, and 21311 phase
(Sr$_2$ScO$_3$FeP and Sr$_2$VO$_3$FeAs )\cite{FeP42622,V42622}. In
the system of (Ba,Sr)$_{1-x}$K$_x$Fe$_2$As$_2$ with the
ThCr$_2$Si$_2$ structure (denoted as 122 phase), the maximum T$_c$
at about 38 K was discovered\cite{Rotter,CWCh} at about x = 0.40.
Large single crystals can be grown in this 122
system.\cite{NiN,Luohq} It has been already found that by
substituting the Fe-sites with the 3d, 4d, and 5d transition metals
like Co\cite{Sefat,XuZA,NiN}, Ni\cite{BaNiFeAs}, Ru\cite{Ru}, Rh, Pd
and Ir\cite{hanfei,NiN2}, the superconductivity can be induced. It
is thus necessary to see whether doping another 5d element Pt can
also induce superconductivity in the 122 phase. In this paper, we
report the successful fabrication of the new superconductor
BaFe$_{2-x}$Pt$_x$As$_2$ with the maximum T$_c$ of about 25 K at the
doping level of x = 0.1. X-ray diffraction (XRD) pattern,
resistivity, synchrotron powder x-ray diffraction, DC magnetic
susceptibility and upper critical field have been measured in this
Pt-doped system. We also explored the phase diagram concerning the
gradual vanishing of the antiferromagnetic order and the
establishment of superconductivity upon doping Pt in this system.

\section{Sample preparation}
The polycrystalline samples BaFe$_{2-x}$Pt$_x$As$_2$ were
synthesized by using a two-step solid state reaction method.
Firstly, BaAs, PtAs and Fe$_{2}$As powders were obtained by the
chemical reaction with Ba pieces, Pt powders (purity 99.95\%), Fe
powders (purity 99.99\%) and As grains. Then they were mixed
together in the formula BaFe$_{2-x}$Pt$_x$As$_2$, ground and pressed
into a pellet shape. All the weighing, mixing and pressing
procedures were performed in a glove box with a protective argon
atmosphere (both H$_2$O and O$_2$ are limited below 0.1 ppm). The
pellet was sealed in a silica tube with 0.2 bar of Ar gas and
followed by a heat treatment at 900 $^o$C for 30 hours. Then it was
cooled down slowly to room temperature.

\section{Experimental data and discussion}
The x-ray diffraction measurements were performed at room
temperature using an MXP18A-HF-type diffractometer with
Cu-K$_{\alpha}$ radiation from 10$^\circ$ to 80$^\circ$ with a step
of 0.01$^\circ$. Synchrotron powder x-ray diffraction (XRD)
experiments were performed on a large Debye-Scherrer camera
installed at SPring-8 beam line BL02B2 by using an imaging plate as
the detector. The wavelength of the x-ray was determined to be 0.602
$\AA$ by using CeO$_2$ as the reference. Glass capillaries with an
inner diameter of 0.3 mm were used to hold the powder samples in
order to eliminate the preferred orientation. The Rietveld
refinements were carried out using GSAS in the angle range of
2$^\circ$ to 75$^\circ$ with an increment of
0.01$^\circ$\cite{GSAS}. The DC magnetization measurements were done
with a superconducting quantum interference device (Quantum Design,
SQUID, MPMS7). The zero-field-cooled magnetization was measured by
cooling the sample at zero field to 2 K, then a magnetic field was
applied and the data were collected during the warming up process.
The field-cooled magnetization data has been collected in the
warming up process after the sample was cooled down to 2 K at a
finite magnetic field. The resistivity measurements were done with a
physical property measurement system PPMS-9T (Quantum Design) with
the four-probe technique. The current direction was reversed for
measuring each point in order to remove the contacting thermal
power.

\subsection{X-ray diffraction}
In order to have a comprehensive understanding to the evolution
induced by the doping effect, we have measured the X-ray diffraction
patterns for BaFe$_{2-x}$Pt$_{x}$As$_{2}$ with x from 0 to 0.25. The
lattice constants along a-axis and c-axis are thus obtained. In
Fig.1 (a), we present the x-ray diffraction patterns of
BaFe$_{2-x}$Pt$_x$As$_2$. It is clear that all main peaks of the
samples can be indexed to the ThCr$_2$Si$_2$ structure. The peaks
marked with asterisks arise from the impurity phase PtAs$_{2}$. As
we can see, only the samples with high doping levels have the
impurity phase PtAs$_{2}$. By fitting the XRD data to the structure
with the software Powder-X,\cite{DongC} we get the lattice constants
of BaFe$_{2-x}$Pt$_{x}$As$_{2}$. In Fig.1 (b)-(c), we show a- and
c-axes lattice parameters for the BaFe$_{2-x}$Pt$_{x}$As$_{2}$
samples. One can see that, by substituting Pt into Fe-sites, the
lattice constant $a$ expands, while $c$ shrinks. This tendency is
similar to the case of doping the Fe-sites with Ni, Rh, Ir or
Ru.\cite{Ru,hanfei,NiN2} For Ni and Co doping, the variation of
a-axis lattice constant seems weaker, but the c-axis lattice
constant drops significantly. Normally a larger a-axis and smaller
c-axis lattice constant would mean that the bond angle of As-Fe-As
is getting larger. Therefore the parameter (a/c)/(a$_{0}$/c$_{0}$)
(a$_{0}$ and c$_{0}$ are the lattice constants for the parent phase)
should tell us some information the doping induced change of bond
angle and superconductivity. In Fig.1(d)-(e) we present the doping
dependence of the unit cell volume and the ratio a/c versus doping
content $x$. Here we compared (a/c)/(a$_{0}$/c$_{0}$) with other
dopants in Fig.1(e).\cite{NiN2,Canfield} The optimal doping for Pt
was found at about 0.1, with (a/c)/(a$_{0}$/c$_{0}$) of about 1.005,
 while it is also near 1.005 for Rh and Pd-doped samples at optimal doping \cite{NiN2}.
 However (a/c)/(a$_{0}$/c$_{0}$) for Co and Ni-doped samples at optimal doping is much smaller than Pt-doped
 samples.\cite{Canfield} Regarding the relatively similar T$_c$
 values in all these systems, it is tempting to conclude that the value of
 (a/c)/(a$_{0}$/c$_{0}$) is not a decisive parameter to govern the
 superconductivity.

\begin{figure}
\includegraphics[width=8cm]{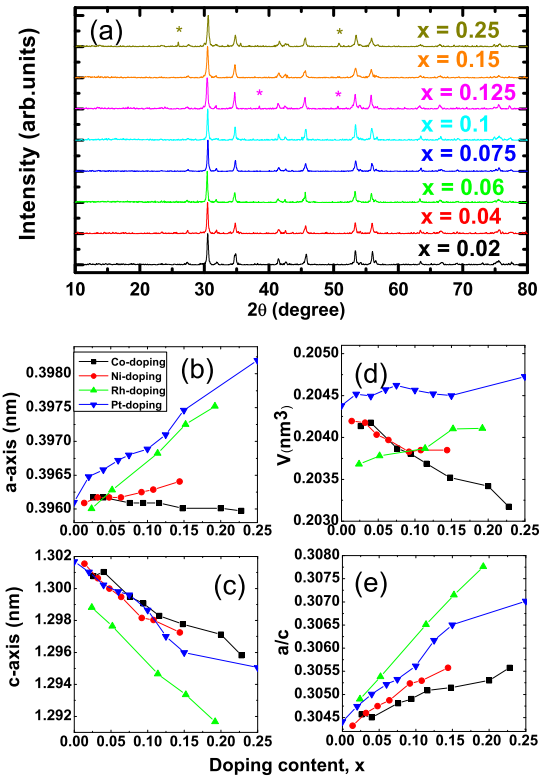}
\caption{(Color online) (a) X-ray diffraction patterns of the
samples BaFe$_{2-x}$Pt$_{x}$As$_{2}$. Almost all main peaks can be
indexed to the tetragonal structure yielding the values of lattice
constants. The asterisks mark the peaks arising from the impurity
phase PtAs$_{2}$. (b)-(c) Doping dependence of the a-axis lattice
constant and c-axis lattice constant of the Co, Ni, Rh and Pt
doping. Here we use the same notation of the doping content $x$ in
BaFe$_{2-x}$TM$_{x}$As$_{2}$ (TM = Co, Ni, Rh and Pt). It is clear
that the a-axis lattice constant expands, while the c-axis one
shrinks monotonically with transition metal substitution. This
systematic evolution clearly indicates that the Pt ions have been
successfully substituted into the Fe-sites. (d) and (e) Doping
dependence of the unit cell volume and the ratio a/c versus doping
content $x$ in BaFe$_{2-x}$TM$_{x}$As$_{2}$ (TM = Co, Ni, Rh and
Pt). The data of Rh and Pt doping are extracted from our
experiments, while those for Ni and Co doping are from the paper of
Canfield.\cite{Canfield}} \label{fig1}
\end{figure}

\subsection{Doping dependence of resistivity}
In Fig.2, we show the temperature dependence of resistivity for
BaFe$_{2-x}$Pt$_{x}$As$_{2}$ samples under zero field in the
temperature region up to 300 K. The resistivity anomaly T$_{an}$ is
determined as the point deviating from the linear part at high
temperatures. As we can see, the parent phase BaFe$_2$As$_{2}$
exhibits a sharp drop of resistivity (resistivity anomaly) at about
140 K, which associates with the formation of the antiferromagnetic
order and structural transition. By doping more Pt, the resistivity
drop was converted to an uprising. We found that the
superconductivity appears in the sample with nominal composition of
x = 0.02, which may be induced by a small amount of superconducting
phase, suggesting slight inhomogeneity in the sample. At this doping
level, the resistivity anomaly T$_{an}$ is about 128 K, being rather
consistent with the structure transition temperature (as shown in
Fig.3). As the doping level was raised to 0.04, the resistivity
anomaly T$_{an}$ moves to 108 K, which is also consistent with the
structure transition temperature. In the sample with x = 0.1, we get
a maximum transition temperature T$_{c}$ of about 25 K, which is
determined by a standard method, i.e., using the crossing point of
the normal state background and the extrapolation of the transition
part with the most steep slope (as shown by the dashed lines in
Fig.5). The transition width determined here with the criterion of
10-90$\%$$\rho_n$ ($\rho_n$ means the normal state resistivity at
the onset transition point) is about 1.87 K. According to Saha et
al.\cite{J. Paglione}, the superconducting transition temperature
T$_{c}$ on the single crystal with x = 0.1 is about 23 K, while the
transition width is about 1.5 K. Comparing with our data on the
polycrystalline sample with x = 0.1, the superconducting transition
temperature T$_{c}$ of our sample is a little higher than theirs,
which may be caused by the slight difference of the doping level.
Meanwhile, the superconducting resistive transition width in our
samples is a little bigger than theirs, which may be caused by the
grain boundaries of polycrystalline samples. However, our data on
the polycrystalline sample in regard to the superconducting
transition behavior is very close to the data on the single crystal
sample.

As mentioned above, there is an impurity phase PtAs$_{2}$ in some
samples, so it's necessary to discuss how the impurity phase might
affect the resistive properties of the material. According to the
X-ray diffraction data, in the low-doped samples, there's no
impurity phase PtAs$_{2}$, therefore here we discuss only the highly
doped samples. In these samples, the impurity scattering caused by
PtAs$_{2}$ may lead to the increase of impurity scattering rate.
However, as we argue below, the sizable impurity scattering effect
in these samples are not induced by the slight amount of PtAs$_{2}$,
but by the intrinsic impurity scattering induced by the dopants at
the Fe-sites. In Fig.2 we show the normalized resistance data. Here,
we take the sample BaFe$_{1.9}$Pt$_{0.1}$As$_{2}$ as an example to
illustrate our point. For this sample, the residual resistance ratio
RRR ($ \equiv$ $\rho$(300K)/$\rho$(30K)) is about 1.35. This value
is very small manifesting a bad metallic behavior and a strong
impurity scattering. However, as evidenced by the X-ray diffraction
data, our sample in this doping is quite clean, which would not lead
to such a strong impurity scattering. Actually, even in the single
crystal samples with the doping to the Fe-sites, RRR is also quite
small. For example, in the single crystals of
BaFe$_{2-x}$M$_{x}$As$_{2}$(M = Pt, Pd, Rh)\cite{NiN2, J. Paglione},
RRR is just around 2. For the single crystals with Ni and Co
doping,\cite{Canfield,LFang} the RRR can at most rises up to 3,
which is much below the value of the
(Sr,Ba)$_{1-x}$K$_x$Fe$_2$As$_2$ which can go up to about 8-10 in
very clean samples.\cite{Luohq} The very large RRR value is
expectable in the samples when the out of plane sites, here such as
the Ba/Sr atoms, are doped with K. So a small RRR in the samples
with doping to the Fe-sites may be a common phenomenon in this
system, indicating an intrinsic strong impurity scattering effect.
It is surprising that the superconductivity can still survive up to
about 25 K even with such a strong impurity scattering effect. This
result challenges the theoretical predictions on the impurity pair
breaking effect with the model of the S$^\pm$ pairing manner in the
iron pnictide superconductors.

\begin{figure}
\includegraphics[width=8cm]{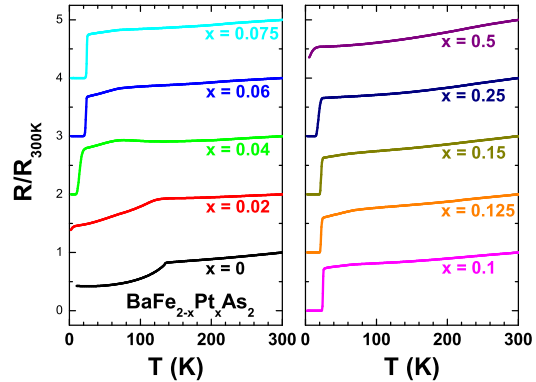}
\caption{(Color online) Temperature dependence of resistivity for
samples BaFe$_{2-x}$Pt$_{x}$As$_{2}$ with x ranging from 0 to 0.5.
The superconductivity starts to appear at x = 0.02, reaching a
maximum T$_c$ of 25 K at about x = 0.1.} \label{fig2}
\end{figure}

\subsection{Synchrotron powder x-ray diffraction}
In order to investigate the relationship between the structural
phase transition and the resistivity anomaly, synchrotron powder
x-ray diffraction (XRD) experiments with the temperature from 10 K
to 300 K were performed on the samples
BaFe$_{1.98}$Pt$_{0.02}$As$_{2}$, BaFe$_{1.96}$Pt$_{0.04}$As$_{2}$,
and BaFe$_{1.9}$Pt$_{0.1}$As$_{2}$. As shown in Fig.3, the refined
crystal structure of BaFe$_{1.98}$Pt$_{0.02}$As$_{2}$ at room
temperature is in good agreement with ThCr$_{2}$Si$_{2}$ structure.
As the temperature decreases, the lattice constants $a$ and $c$
shrink a bit. The inset of Fig.3.(a) shows the (213) Bragg
reflection peak as a function of temperature. The splitting of that
peak indicates that the sample undergoes a tetragonal to
orthorhombic distortion. The space group symmetry changes from
tetragonal (I4=mmm) to orthorhombic (Fmmm) at about 125 K. As
mentioned above, the resistivity anomaly T$_{an}$ of
BaFe$_{1.98}$Pt$_{0.02}$As$_{2}$ is about 128 K, which is very close
to that determined from the synchrotron data. So the resistivity
anomaly T$_{an}$ is rather consistent with structural transition
temperature at the doping level x = 0.02. At the doping level x =
0.04, T$_{an}$ is about 108 K, while the structural transition
occurs between 100 K and 120 K (not shown here). At the doping level
x = 0.1, no structural transition was found in the synchrotron
experiment at all temperatures.
\begin{figure}
\includegraphics[width=8cm]{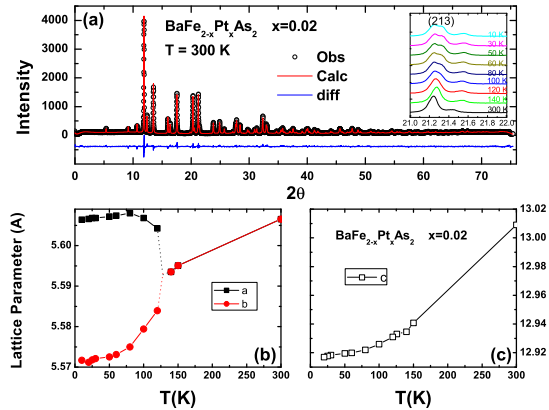}
\caption {(Color online) Synchrotron powder x-ray diffraction and
the Rietveld fit for BaFe$_{1.98}$Pt$_{0.02}$As$_{2}$. The lattice
parameters shrink with the decreasing of temperature. At about 125
K, the tetragonal-orthorhombic structural phase transition occurs,
this can be easily seen in the inset shown for the peak (213). }
\label{fig3}
\end{figure}

\subsection{The electronic phase diagram}
\begin{figure}
\includegraphics[width=8cm]{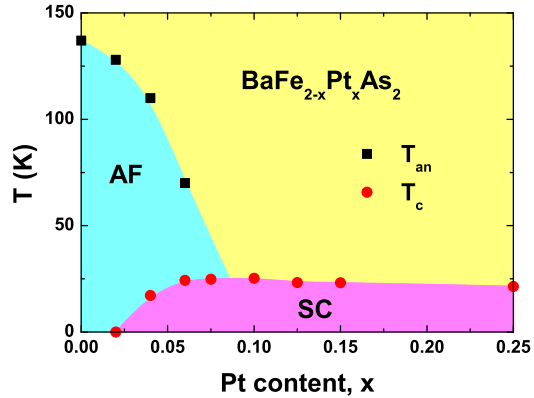}
\caption {(Color online) Phase diagram of
BaFe$_{2-x}$Pt$_{x}$As$_{2}$ within the doping range of x = 0 to
0.25. The temperature of resistivity anomaly represents the starting
point of the upturning of resistivity, i.e., the deviating point
from a rough T-linear behavior in the high temperature region. The
superconductivity starts to appear at about x = 0.02, the T$_c$
value reaches a maximum of 25 K at x = 0.1. The long-tail like
overdoped region may be induced by the slow degradation of the AF
spin fluctuations, or the percolative superconductivity.}
\label{fig4}
\end{figure}
Based on the measurements described above, a phase diagram of
BaFe$_{2-x}$Pt$_{x}$As$_{2}$ within the doping range of x from 0 to
0.25 was given in Fig.4. The T$_{an}$ was defined as the
temperatures of the anomaly in resistivity, and T$_{c}$ was
determined from the onset of superconducting resistive transition.
As we can see, with increasing Pt content, the temperature of the
resistivity anomaly which corresponds to the tetragonal-orthorhombic
structural / antiferromagnetic transition is driven down, and the
superconductivity emerges at x = 0.02, reaching a maximum T$_c$ of
25 K at x = 0.1. This general phase diagram looks very similar to
that with Ni and Pd doping.\cite{XuZA,Canfield} Since Pt locates
just below Ni and Pd in the periodic table of elements, we would
conclude that the superconductivity induced by Pt doping shares the
similarity as that of Ni doping.

One of the interesting point here is that, the superconductivity
transition at about 23 K was observed even up to the doping level of
x = 0.25, making the superconducting phase region extremely
asymmetric. This can be naturally understood as the slow degradation
of the AF spin fluctuations in the overdoped region if assuming that
the paring is through exchanging the AF spin fluctuations. As
indicated in our previous work in the Co-doped Ba-122 single
crystals,\cite{LFang} the superconducting "dome" is not symmetric at
all in the iron pnictide superconductors. In the underdoped region,
the superconducting transition temperature ramps quickly up to the
maximum value since the superconducting phase wins more and more
density of states from the AF order. It is these quasiparticles that
pair via exchanging the residual AF spin fluctuation and form the
superconducting condensate. While, in the overdoped region, the
whole Fermi surfaces will involve in the superconducting pairing,
but now the pairing strength which is governed by the AF spin
fluctuation becomes weaker and weaker. The extended superconducting
transition at such a high doping level may also be partly attributed
to the percolative superconductivity. The doping may have an upper
solubility limit. Above this doping level, the system will
chemically phase-separate into two different regions, one with
superconductivity at almost the optimized T$_c$, while the others
are the impurity phases, such as PtAs$_2$. Further experiments on
single crystals may decide whether this long-tail like over doped
region is an intrinsic feature for the Pt doped samples.

\subsection{Upper critical field}
\begin{figure}
\includegraphics[width=8cm]{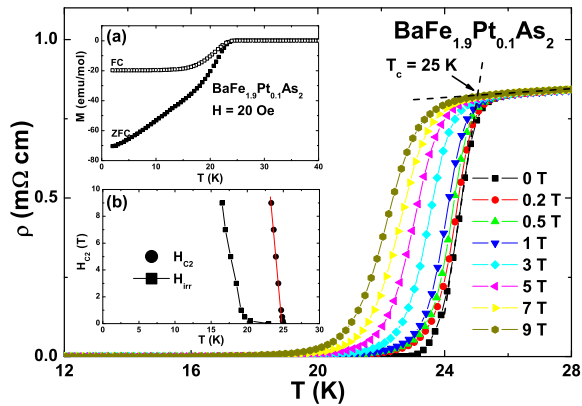}
\caption {(Color online) Temperature dependence of resistivity for
the sample BaFe$_{1.9}$Pt$_{0.1}$As$_{2}$ at different magnetic
fields. The dashed line indicates the extrapolated resistivity in
the normal state. One can see that the superconductivity seems to be
robust against the magnetic field and shifts slowly to lower
temperatures. The inset(a) shows the dc susceptibility data using
the zero-field-cooling and field-cooling modes with a dc magnetic
field of 20 Oe. The inset(b) gives the upper critical field
determined using the criterion of 90\%$\rho_n$. A slope of
-dH$_{c2}$/dT = 5.4T/K near T$_c$ is found here. The irreversibility
line H$_{irr}$ taking with the criterion of 0.1\% $\rho_n$ is also
presented in the inset.} \label{fig5}
\end{figure}

In Fig.5, we present the temperature dependence of resistivity under
different magnetic fields for the sample with x = 0.1. As shown in
the inset(a) of Fig.5, just as many other iron pnictide
superconductors, the diamagnetic signal is very strong. However,
from the zero-field-cooling M(T) curve one can also see that the low
temperature part is not flat. This strong temperature dependence was
not seen in the field-cooling M(T) data. We explain this strong
temperature dependent ZFC M(T) curve as due to the decaying of the
Meissner screening current, which is induced by the easy motion the
of magnetic flux (even not the superconducting quantized flux lines)
through the weak-links at the grain boundaries. This happens quite
often in the polycrystalline samples. We thus used the criterion of
$90\%\rho_n$ to determine the upper critical field and show the data
in the inset(b) of Fig.5. A slope of -dH$_{c2}$/dT = 5.4 T/K can be
obtained here. This is a rather large value which indicates a rather
high upper critical field in this system. By using the
Werthamer-Helfand-Hohenberg (WHH) formula\cite{WHH} for a single
band system $H_{\mathrm{c}2}(0)=-0.69(dH_{c2}/dT)|_{T_c}T_c$, the
value of zero temperature upper critical field can be estimated.
Taking $T_\mathrm{c}= 25\;$K, we can get $H_{\mathrm{c}2}(0) \approx
93 T$ roughly. This is a very large upper critical field, just as in
K-doped\cite{WangZSPRB} and Co-doped samples\cite{Jo}. However, the
high upper critical field is just an estimate of the data based on
the single band model with the assumption that the upper critical
field is determined by the orbital pair breaking effect. In the case
of multi-band superconductivity and paramagnetic limit for the upper
critical field, the value of H$_{c2}(0)$ will differ from the value
predicted by the WHH theory. Therefore an accurate determination of
H$_{c2}(0)$ needs to measure the resistive transitions directly
under high magnetic fields. While the high value of -dH$_{c2}$/dT at
T$_c$ at least indicates a rather strong pairing strength with a
rather high quasiparticle density of states.

A very interesting point uncovered by our experiment is that the
maximum T$_c$ by doping Pt in BaFe$_2$As$_2$ is 25 K, which is close
to that by doping Ni, Co and Ir at the Fe sites. Manifold interests
can be raised here. (1) Although the ionic sizes are rather
different among them, while the maximum T$_c$ is not influenced by
them, which may indicate that the non-magnetic centers paly a
trivial role in the pair breaking effect. (2) Although the mass of
the 5d element Pt is much heavier than the mass of the 3d elements,
such as Ni, again the maximum T$_c$ is the same by doping them to
the Fe sites, which trivializes the importance of phonon mediated
mechanism in the occurrence of superconductivity. (3) Normally the
5d transition elements have much wider band and stronger spin
orbital coupling effect, while the similar maximum T$_c$ may suggest
that the electron itineracy as well as the spin-orbital coupling are
not the determining factors for the occurrence of superconductivity.
Our results here shed new lights in the understanding of the
mechanism of superconductivity in the iron pnictide superconductors.

\section{Conclusions}

In summary, superconductivity has been found in
BaFe$_{2-x}$Pt$_{x}$As$_2$ with the maximum T$_c$ = 25 K. The phase
diagram obtained here is quite similar to that by doping Co, Ni, Rh
and Ir to the Fe sites. The resistivity anomaly temperature T$_{an}$
is rather consistent with that of the structure transition. It is
found that all samples with doping to the Fe sites showed an very
small RRR (below 3), which is much smaller than the value (up to 10)
with the out-off plane dopants, like K-doped (Ba,Sr)-122. This
indicates the doping at Fe-sites naturally leads to a strong
intrinsic impurity scattering. While surprisingly, superconductivity
can still survive up to about 25 K, even with such a strong impurity
scattering effect. The superconductivity is rather robust against
the magnetic field with a slope of -dH$_{c2}$/dT = 5.4 T/K near
T$_c$ at the doping level x = 0.1. Our results clearly indicate that
the superconductivity can also be easily induced in BaFe$_2$As$_2$
by replacing Fe with Pt. This discovery may trivialize the phonon
effect, electron itinerancy, and the spin-orbital coupling in the
occurrence of superconductivity.

Note added: When preparing this manuscript, we became aware that J.
Paglione et al posted a paper concerning the superconductivity in
Pt-doped BaFe$_2$As$_2$.\cite{J. Paglione} Their results are
consistent with ours although they report the result for only one
doping level.

This work was supported by the Natural Science Foundation of China,
the Ministry of Science and Technology of China (973 Projects
No.2006CB601000, No. 2006CB921802), and Chinese Academy of Sciences
(Project ITSNEM).

\end{document}